\documentclass[]{article}
\usepackage{amsfonts}
\usepackage{graphicx}
\usepackage[paperwidth=8.5in, paperheight=11in]{geometry}
\usepackage{float} 
\usepackage[section]{placeins}
\begin{document}
\newcommand{\R}{\mathbb{R}}
\newcommand{\C}{\mathbb{C}}
\centerline{\bf NEWMAN-JANIS ALGORITHM REVISITED}

\

\

\centerline{O. Brauer and H. A. Camargo}
\centerline{\it Facultad de Ciencias, Universidad Nacional Aut\'onoma de M\'exico}
\centerline{\it Circuito Exterior, Ciudad Universitaria, 04510, M\'exico D. F., M\'exico} 
\centerline{and}
\centerline{M. Socolovsky}
\centerline{\it  Instituto de Ciencias Nucleares, Universidad Nacional Aut\'onoma de M\'exico}
\centerline{\it Circuito Exterior, Ciudad Universitaria, 04510, M\'exico D. F., M\'exico} 

\

{\it The purpose of the present article is to show that the Newman-Janis and Newman et al algorithm used to derive the Kerr and Kerr-Newman metrics respectively, automatically leads to the extension of the initial non negative polar radial coordinate $r$ to a cartesian coordinate $r^\prime$ running from $-\infty$ to $+\infty$, thus introducing in a natural way the region $-\infty<r^\prime<0$ in the above spacetimes. Using Boyer-Lindquist and ellipsoidal coordinates, we discuss some geometrical aspects of the positive and negative regions of $r^\prime$, like horizons, ergosurfaces, and foliation structures.}

\

{\bf 1. Introduction}

\

The uncharged Kerr ($K$) (Kerr, 1963) and the charged Kerr-Newman ($KN$) (Newman et al, 1965) axially symmetic stationary spacetimes have, in contradistinction to the charged static spherical Reissner-N\"{o}rdstrom ($RN$) solution (Reissner, 1916; N\"{o}rdstrom, 1918), and the chargeless static spherical Schwarzschild solution (Schwarzschild, 1916), an asymptotically flat region which involves, both in the Eddington-Finkelstein ($EF$) (Eddington, 1924; Finkelstein, 1958) and  Boyer-Lindquist ($BL$) (Boyer and Lindquist, 1967) coordinates, a radial cordinate taking not only positive and zero values, but also negative ones, form 0 to -$\infty$. This strange situation is usually explained by the fact that, since the curvature singularity of both the $K$ and $KN$ solutions is in the circular boundary of a non singular open disk in the equatorial plane, the spacetimes can be continued through it, to regions in which the radial coordinate becomes negative (regions $III$ and $III^\prime$ in the Penrose-Carter diagram (Penrose, 1963; Carter, 1966), section {\bf 8}). As it stands, the argument, though correct, does not imply the necessity of this continuation, but only it allows for its possibility. 

\

In this note we show that the complexification procedure involved in the Newman-Janis algorithm ($NJA$) (Newman and Janis, 1965) used to derive the $K$ and $KN$ metrics, automatically leads to the extension to negative values of the originally non negative polar radial coordinate, leaving from the outset with coordinates $(u^\prime,r^\prime,\theta,\varphi)$ (or $(t^\prime,r^\prime, \theta,\phi)$) taking values in $\R^2\times S^2$ i.e. two cartesian ($u^\prime,r^\prime$ or $t^\prime,r^\prime$) and two compact ($\theta,\varphi$ or $\theta,\phi$) coordinates ($u^\prime$ is the retarded $EF$ time and $t$ the $BL$ time). Though at the epoque of its inception and for many years, the Newman-Janis and Newman et al derivations respectively of the $K$ and $KN$ metrics were considered as flukes, recent work by Drake and Szekeres (Drake and Szekeres, 2000) has put the algorithm on a more solid ground by proving uniqueness theorems for the kind of solutions which can be derived using the algorithm. 

\

In sections {\bf 2} to {\bf 5} we review the $NJA$ derivation of the $K$ and $KN$ metrics emphasizing, in section {\bf 4}, how the complexification of the radial coordinate automatically implies the range $(-\infty,+\infty)$ for its real part $r^\prime$. Using $BL$ (section {\bf 6}) and ellipsoidal coordinates, in section {\bf 7} we exhibit the horizons, ergosurfaces, and foliation structures of the $KN$ and $K$ solutions (respectively for the cases $M^2>a^2+Q^2$ and $M^2>a^2$) in both regions $r^\prime >0$ and $r^\prime <0$, and give a brief discussion of the spatial topology of these constructions. For completeness, the basic cell of the Penrose-Carter diagram of the $K$ and $KN$ spacetimes is exhibited in section {\bf 8}.  

\

{\bf 2. Reissner-N\"{o}rdstrom spacetime} 

\

Our starting point is the Reissner-N\"{o}rdstrom ($RN$) spacetime written in terms of the retarded Eddington
-Finkelstein coordinates $(u,r,\theta,\varphi)$ with $u\in\R=(-\infty,+\infty)$, $r\in\R_{\geq}=[0,+\infty)$, and $\theta,\varphi\in S^2$ i.e. $\theta\in[0,\pi]$ and $\varphi\in [0,2\pi)$: $$ds^2_{RN}=fdu^2+2dudr-r^2d^2\Omega, \ \ d^2\Omega=d\theta^2+sin^2\theta d\varphi^2 \eqno{(1)}$$ with $$f={{r^2-2Mr+Q^2}\over{r^2}}=1-{{2M}\over{r}}+{{Q^2}\over{r^2}}, \ \ [f]=[L]^0, \eqno{(2)}$$ where $M$ is the gravitating (positive) mass and $Q^2=q^2+p^2$ where $q$ is the electric charge and $p$ is the hypotetical abelian (Dirac) magnetic charge. The metric corresponding to (1) is $${g_{\mu\nu}}_{RN}=\pmatrix{f&1&0&0\cr 1&0&0&0\cr 0&0&-r^2&0\cr 0&0&0& -r^2sin^2\theta\cr}, \ \ det {g_{\mu\nu}}_{RN}=-r^4sin^2\theta,\eqno{(3)}$$ with inverse $${g^{\mu\nu}}_{RN}=\pmatrix{0&1&0&0\cr 1&-f&0&0\cr 0&0&-r^{-2}&0\cr 0&0&0&-r^{-2}sin^{-2}\theta\cr}.\eqno{(4)}$$ ($\mu=0,1,2,3$ respectively correspond to $u,r,\theta,\varphi.$)

\

{\bf 3. Null tetrad}

\

At each point of the $RN$ manifold we can choose as a basis of the corresponding tangent space the null tetrad consisting of the following linear independent 4-vectors:

\

$$l=l^\mu{{\partial}\over{\partial x^\mu}}=\delta^\mu_1{{\partial}\over{\partial x^\mu}}={{\partial}\over{\partial x^1}}={{\partial}\over{\partial r}},\eqno{(5a)}$$ $$n=n^\mu{{\partial}\over{\partial x^\mu}}=(\delta^\mu_0-{{f}\over{2}}\delta^\mu_1){{\partial}\over{\partial x^\mu}}={{\partial}\over{\partial u}}-{{f}\over{2}}{{\partial}\over{\partial r}},\eqno{(5b)}$$ $$m=m^\mu{{\partial}\over{\partial x^\mu}}={{1}\over{\sqrt{2}r}}(\delta^\mu_2+{{i}\over{sin\theta}}\delta^\mu_3){{\partial}\over{\partial x^\mu}}={{1}\over{\sqrt{2}r}}({{\partial}\over{\partial\theta}}+{{i}\over{sin\theta}}{{\partial}\over{\partial\varphi}}),\eqno{(5c)}$$ $$\bar{m}=\bar{m}^\mu{{\partial}\over{\partial x^\mu}}={{1}\over{\sqrt{2}r}}(\delta^\mu_2-{{i}\over{sin\theta}}\delta^\mu_3){{\partial}\over{\partial x^\mu}}={{1}\over{\sqrt{2}r}}({{\partial}\over{\partial\theta}}-{{i}\over{sin\theta}}{{\partial}\over{\partial\varphi}}),\eqno{(5d)}$$ i.e. $$l^\mu=(l^u,l^r,l^\theta,l^\varphi)=(0,1,0,0),\eqno{(5a')}$$ $$n^\mu=(n^u,n^r,n^\theta,n^\varphi)=(1,-{{f}\over{2}},0,0),\eqno{(5b')}$$ $$m^\mu=(m^u,m^r,m^\theta,m^\varphi)={{1}\over{\sqrt{2}r}}(0,0,1,{{i}\over{sin\theta}}),\eqno{(5c')}$$ $$\bar{m}^\mu=(\bar{m}^u,\bar{m}^r,\bar{m}^\theta,\bar{m}^\varphi)={{1}\over{\sqrt{2}r}}(0,0,1,-{{i}\over{sin\theta}}),\eqno{(5d')}$$ with covariant components $b_\mu={g_{\mu\nu}}_{RN}b^\nu$ given by $$l_\mu=(1,0,0,0,),\eqno{(5a'')}$$ $$n_\mu=({{f}\over{2}},1,0,0),\eqno{(5b'')}$$ $$m_\mu={{1}\over{\sqrt{2}r}}(0,0,-r^2,-ir^2sin\theta),\eqno{(5c'')}$$ $$\bar{m}_\mu={{1}\over{\sqrt{2}r}}(0,0,-r^2,ir^2sin\theta).\eqno{(5d'')}$$ The scalar products $a\cdot b={g_{\mu\nu}}_{RN}a^\mu b^\nu$ of the members of the tetrad are given in the following table: $$\matrix{a\cdot b&l&n&m&\bar{m}\cr l&0&1&0&0\cr n&1&0&0&0\cr m&0&0&0&-1\cr \bar{m}&0&0&-1&0\cr}$$

\

\centerline{Table I}

\

It is an easy exercise to verify that the quantity $$\tilde{g}^{\mu\nu}=(l^\mu n^\nu+l^\nu n^\mu)-(m^\mu\bar{m}^\nu+m^\nu\bar{m}^\mu)\eqno{(6)}$$ is nothing but the inverse $RN$ metric: $$\tilde{g}^{\mu\nu}={g^{\mu\nu}}_{RN}.\eqno{(7)}$$ It is interesting to note that if the tetrad is denoted by $$(e_1,e_2,e_3,e_4)=(l,n,m,\bar{m}),\eqno{(8)}$$ then $$\tilde{g}^{\mu\nu}={e_a}^\mu J^{ab}{e_b}^\nu\eqno{(9)}$$ with $$J=\pmatrix{\sigma_1&0\cr 0&-\sigma_1\cr},\eqno{(10)}$$ where $\sigma_1=\pmatrix{0&1\cr 1&0\cr}$ is the Pauli matrix. $J$ is related to the $\alpha^k=\pmatrix{0&\sigma_k\cr \sigma_k&0\cr}$, $k=1,2,3$, matrices of the standard representation of the Dirac equation, with $\sigma_2=\pmatrix{0&-i\cr i&0}$ and $\sigma_3=\pmatrix{1&0\cr 0&-1\cr}$, through the similarity transformations $$S_k^{-1}JS_k=\alpha^k, \ S_1=\pmatrix{1&1\cr 1&-1\cr}, \ S_2=\pmatrix{1&\sigma_1\sigma_2\cr 1&-\sigma_1\sigma_2\cr}, \ S_3=\pmatrix{1&\sigma_1\sigma_3\cr 1&-\sigma_1\sigma_3},\eqno{(11)}$$ with $$S_1^{-1}={{1}\over{2}}\pmatrix{1&1\cr 1&-1}, \ S_2^{-1}={{1}\over{2}}\pmatrix{1&1\cr -i\sigma_3&i\sigma_3\cr}, \ S_3^{-1}={{1}\over{2}}\pmatrix{1&1\cr i\sigma_2&-i\sigma_2}.\eqno{(12)}$$ It is curious that a relation with the Pauli and Dirac matrices appears here, since the $RN$ solution is static without rotation. The next complexification and appearance of the $KN$ metric does not change the situation, since the $J$ matrix tacitly involved in eq. (22) is the same as that in eq. (9). We leave this apparent ``accident" opened to further research.

\

{\bf 4. Complexification}

\

The following step is the crucial element of the NJA: {\it the coordinates} $r$ {\it and} $u$ {\it are complexified} and a new real positive parameter $a$ (later identified with the angular momentum per unit mass) is introduced:

\

$$r\in\R_{\geq}\longrightarrow r\in \C, \ \ r=r^\prime-iacos\theta,\eqno{(13)}$$ $$u\in\R \longrightarrow u\in\C, \ \ u=u^\prime+iacos\theta, \eqno{(14)}$$ with $\bar{r}=r^\prime+iacos\theta$ and $\bar{u}=u^\prime-iacos\theta$. But now $$r^\prime\in(-\infty,+\infty), \eqno{(15)}$$ i.e. $r^\prime$ has become a {\it cartesian coordinate}. In particular, this will imply that when dealing with the Boyer-Lindquist coordinates in section {\bf 6}, no appeal for an analytic continuation to the asymptotically flat (A.F.) region $r^\prime<0$ through the open disk $y^2+z^2<a^2$ where no singularity occurs will be required: implicitly, this analytic continuation was already done through the complexification (13). The domains of definition of $r$ and $u$ are shown in Figure 1. Clearly, $[a]=[L]^1$. 

\

\begin{figure}[H] 
\includegraphics[width=\linewidth]{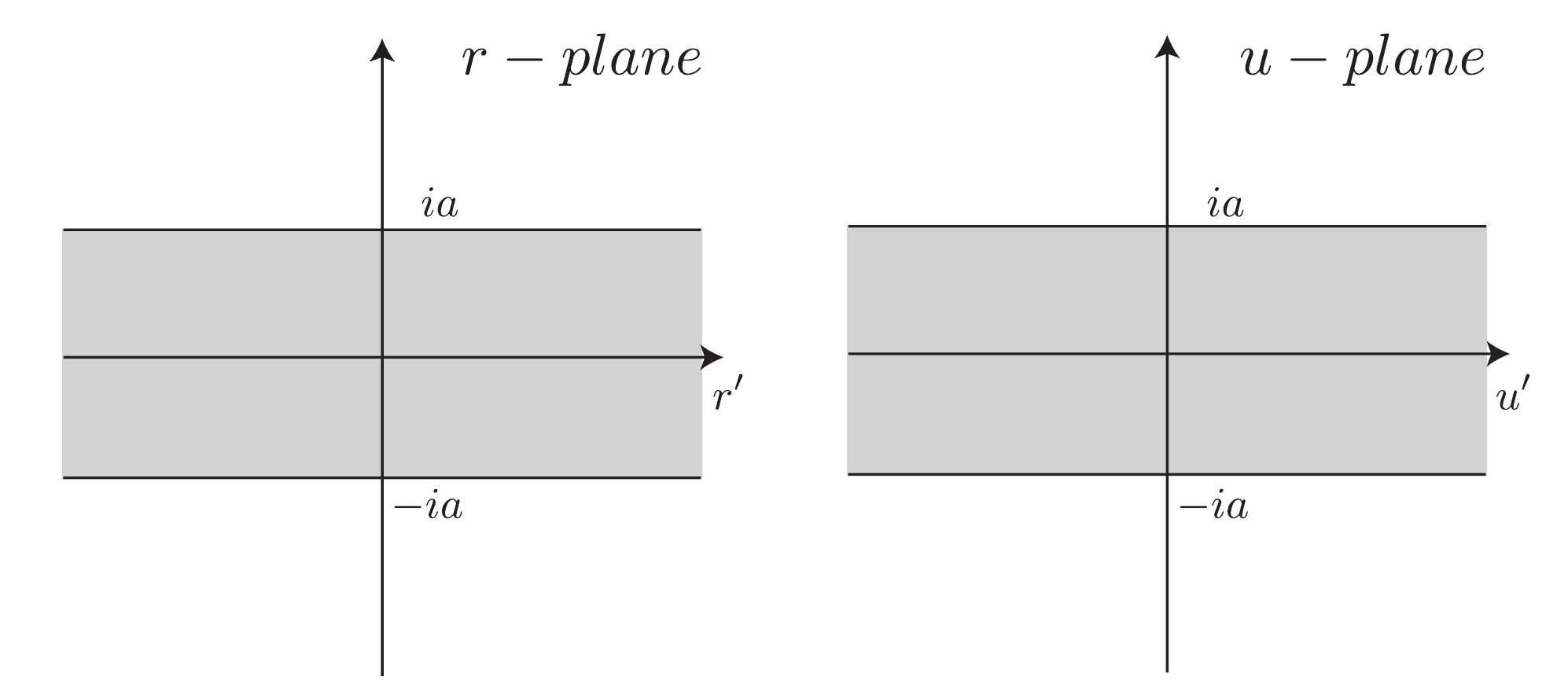}
\caption{Domains of definition of the coordinates $r$ and $u$ after the complexification (13).}
\end{figure}
\

The new set of real coordinates is $(u^\prime,r^\prime,\theta^\prime,\varphi^\prime)$ with $u^\prime\in(-\infty,+\infty)$, $\theta^\prime=\theta$, and $\varphi^\prime=\varphi$. The total domain of these coordinates is $\R^2\times S^2$.

\

Since in the end one needs a real spacetime, $f$ must remain real and so its change is given by $$f(r)\longrightarrow f(r,\bar{r})=1-M({{1}\over{r}}+{{1}\over{\bar{r}}})+{{Q^2}\over{r\bar{r}}}=1-{{2Mr^\prime-Q^2}\over{\Sigma}}\eqno{(16)}$$ where $$\Sigma={r^\prime}^2+a^2cos^2\theta.\eqno{(17)}$$ ($[\Sigma]=[L]^2$.) The transformation of the tetrad is $${e_a}^\mu\longrightarrow {e_a^\prime}^{\mu}={{\partial{x^\prime}^\mu}\over{\partial x^\nu}}{e_a}^\nu\equiv ({l^\prime}^\mu,{n^\prime}^\mu,{m^\prime}^\mu,{\bar{m^\prime}}^\mu)\eqno{(18)}$$ with $$\pmatrix{{{\partial u^\prime}\over{\partial u}} & {{\partial u^\prime}\over{\partial r}} & {{\partial u^\prime}\over{\partial \theta}} & {{\partial u^\prime}\over{\partial \varphi}} \cr {{\partial r^\prime}\over{\partial u}} & {{\partial r^\prime}\over{\partial r}} & {{\partial r^\prime}\over{\partial \theta}} & {{\partial r^\prime}\over{\partial \varphi}} \cr {{\partial \theta^\prime}\over{\partial u}} & {{\partial \theta^\prime}\over{\partial r}} & {{\partial \theta^\prime}\over{\partial \theta}} & {{\partial \theta^\prime}\over{\partial \varphi}} \cr {{\partial \varphi^\prime}\over{\partial u}} & {{\partial \varphi^\prime}\over{\partial r}} & {{\partial \varphi^\prime}\over{\partial \theta}} & {{\partial \varphi^\prime}\over{\partial \varphi}}}=\pmatrix{1&0&iasin\theta&0\cr 0&1&-iasin\theta&0\cr 0&0&1&0\cr 0&0&0&1\cr}\eqno{(19)}$$ and $$l^\nu=\delta^\nu_1, \ n^\nu=\delta^\nu_0-{{1}\over{2}}f(r,\bar{r})\delta^\nu_1, \ m^\nu={{1}\over{\sqrt{2}\bar{r}}}(\delta^\nu_2+{{i}\over{sin\theta}}\delta^\nu_3), \ \bar{m^\nu}={{1}\over{\sqrt{2}r}}(\delta^\nu_2-{{i}\over{sin\theta}}\delta^\nu_3).\eqno{(20)}$$ One obtains: $${l^\prime}^\mu=\delta^\mu_1, \eqno{(21a)}$$ $${n^\prime}^\mu=\delta^\mu_0-{{1}\over{2}}(1-{{2Mr^\prime-Q^2}\over{\Sigma}})\delta^\mu_1, \eqno{(21b)}$$ $${m^\prime}^\mu={{1}\over{\sqrt{2}(r^\prime+iacos\theta)}}((\delta^\mu_0-\delta^\mu_1)iasin\theta+\delta^\mu_2+\delta^\mu_3{{i}\over{sin\theta}}),\eqno{(21c)}$$ $${\bar{m^\prime}}^\mu={{1}\over{\sqrt{2}(r^\prime-iacos\theta)}}(-(\delta^\mu_0-\delta^\mu_1)iasin\theta+\delta^\mu_2-\delta^\mu_3{{i}\over{sin\theta}}).\eqno{(21d)}$$

\

{\bf 5. Kerr-Newman metric}

\

The ``miracle", though in some way justified by Drake and Szekeres through the proof of important uniqueness theorems from the NJA for vacuum and Einstein-Maxwell solutions, is that the quantity $${g^\prime}^{\mu\nu}=({l^\prime}^\mu{n^\prime}^\nu+{l^\prime}^\nu{n^\prime}^\mu)-({m^\prime}^\mu\bar{{m^\prime}}^\nu+{m^\prime}^\nu\bar{{m^\prime}}^\mu)\eqno{(22)}$$ is the inverse Kerr-Newman ($KN$) metric in Eddington-Finkelstein retarded coordinates $(u^\prime,r^\prime,\theta,\varphi)$: $${g^\prime}^{\mu\nu}={g^{\mu\nu}}_{KN}. \eqno{(23)}$$ In fact, a straightforward calculation leads to $${g^\prime}^{\mu\nu}=\pmatrix{g^{u^\prime u^\prime}&g^{u^\prime r^\prime}&g^{u^\prime \theta}&g^{u^\prime \varphi}\cr \cdot&g^{r^\prime r^\prime}&g^{r^\prime \theta}&g^{r^\prime \varphi}\cr \cdot&\cdot&g^{\theta \theta}&g^{\theta \varphi}\cr \cdot&\cdot&\cdot&g^{\varphi\varphi}\cr}=$$
$$\pmatrix{{{-a^2sin^2\theta}\over{\Sigma}}&{{{r^\prime}^2+a^2}\over{\Sigma}}&0&-{{a}\over{\Sigma}}\cr \cdot&-{{\Sigma-2Mr^\prime+Q^2+a^2sin^2\theta}\over{\Sigma}}&0&{{a}\over{\Sigma}}\cr \cdot&\cdot&-{{1}\over{\Sigma}}&0\cr \cdot&\cdot&\cdot&-{{1}\over{\Sigma sin^2\theta}}}\eqno{(24)}$$ with inverse $${g_{\mu\nu}}_{KN}=\pmatrix{1-{{2Mr^\prime-Q^2}\over{\Sigma}}&1&0&asin^2\theta{{2Mr^\prime-Q^2}\over{\Sigma}}\cr \cdot&0&0&-asin^2\theta\cr \cdot&\cdot&-\Sigma&0\cr \cdot&\cdot&\cdot&-sin^2\theta{{A}\over{\Sigma}}\cr}\eqno{(25)}$$ with $$A=\Sigma({r^\prime}^2+a^2)+a^2sin^2\theta{{2Mr^\prime-Q^2}\over{\Sigma}}=({r^\prime}^2+a^2)^2-a^2sin^2\theta\Delta,\eqno{(26)}$$ and $$\Delta={r^\prime}^2+a^2-2Mr^\prime+Q^2.\eqno{(27)}$$ The square of the spacetime element is $$ds^2_{KN}=$$
$$(1-{{2Mr^\prime-Q^2}\over{\Sigma}})d{u^\prime}^2+2du^\prime dr^\prime+2asin^2\theta{{2Mr^\prime-Q^2}\over{\Sigma}}du^\prime d\varphi-2asin^2\theta dr^\prime d\varphi-\Sigma d\theta^2-sin^2\theta{{A}\over{\Sigma}}d\varphi^2.\eqno{(28)}$$ $ds^2_{KN}$ reduces to $ds^2_{RN}$ for $a=0$ and $r^\prime\geq 0$. 

\

{\bf 6. Boyer-Lindquist coordinates}

\

The change of coordinates
 $$dt=du^\prime-{{{r^\prime}^2+a^2}\over{\Delta}}dr^\prime, \ \ d\phi=d\varphi-{{a}\over{\Delta}}dr^\prime \eqno{(29)}$$ 
 leads to the Boyer-Lindquist form of the Kerr-Newman spacetime:
  $$ds^2_{KN}\vert_{BL}={{\Delta-a^2sin^2\theta}\over{\Sigma}}dt^2-{{\Sigma}\over{\Delta}}d{r^\prime}^2-\Sigma d\theta^2-sin^2\theta{{A}\over{\Sigma}}d\phi^2+{{2asin^2\theta}\over{\Sigma}}({r^\prime}^2+a^2-\Delta)dtd\phi\eqno{(30)}$$ i.e. $${g_{\mu\nu}}_{KN}\vert_{BL}=\pmatrix{1-{{2Mr^\prime-Q^2}\over{\Sigma}}&0&0&{{asin^2\theta}\over{\Sigma}}({r^\prime}^2+a^2-\Delta)\cr \cdot&-{{\Sigma}\over{\Delta}}&0&0\cr \cdot&\cdot&-\Sigma&0\cr \cdot&\cdot&\cdot&-sin^2\theta{{A}\over{\Sigma}}\cr}.\eqno{(31)}$$
   Horizons $H_+$ and $H_-$ are defined by the zeros of $\Delta$; it is clear that only for $r^\prime >0$ and $M^2\geq a^2+Q^2$ horizons exist, with $$r^\prime_\pm=M\pm\sqrt{M^2-(a^2+Q^2)}.\eqno{(32)}$$ It can be easily seen that for $M^2>a^2+Q^2$, $r^\prime_- <\sqrt{a^2+Q^2}$, in particular $r^\prime_-<a$ for the Kerr case $Q^2=0$; for the extreme cases $M^2=a^2+Q^2$, $r^\prime_-=r^\prime_+=M=\sqrt{a^2+Q^2}$. 

\

For $r^\prime < 0$, $$\Delta={r^\prime}^2+a^2+2M\vert r^\prime\vert+Q^2>0,\eqno{(33)}$$ which has no real roots. The same occurs for the ergosurfaces $S_+$ and $S_-$ whose equations are given by the zeros of ${g_{tt}}_{KN}\vert_{BL}$; for $r^\prime < 0$, $${g_{tt}}_{KN}\vert_{BL}=1+{{2M\vert r^\prime\vert +Q^2}\over{\Sigma}}>1.\eqno{(34)}$$ Also, as is well known, from (17), (26), (27) and (30), $$ds^2_{KN}\vert_{BL}\longrightarrow dt^2-d{r^\prime}^2-{r^\prime}^2(d\theta^2+sin^2\theta d\phi^2)\eqno{(35)}$$ as $r^\prime\to \pm\infty$ i.e. the metric is A.F. in both the $r^\prime>0$ and $r^\prime<0$ regions.

\

{\bf 7. Generalized ellipsoidal coordinates} 

\

As is well known, the use of Kerr-Schild ($KS$) coordinates for the $KN$ metric and their restriction to ellipsoidal coordinates ($EC$) for the Kerr ($K$) metric ($ds^2_K=ds^2_{KN}\vert_{Q^2=0}$), allows to show that both metrics are flat (Minkowskian) for $M^2=Q^2=0$ in the $KN$ case and $M^2=0$ in the $K$ case. The general form of the spatial part of these coordinates allowing for both positive and negative values of $r^\prime$ is $$x_\pm=\sqrt{{r^\prime}^2+a^2} \ sin\theta \ cos(\phi+F(r^\prime)),\eqno {(36a)}$$ $$y_\pm=\sqrt{{r^\prime}^2+a^2} \ sin\theta \ sin(\phi+F(r^\prime)),\eqno {(36b)}$$ $$z_\pm=r^\prime cos\theta \eqno{(36c)}$$ with the + (-) sign corresponding to $r^\prime>0$ ($r^\prime<0$) and $$F(r^\prime)=\lbrace\matrix{-arc \ tan({{a}\over{r^\prime}}), \ KS \cr 0, \ EC}. \eqno{(37)}$$ $x_\pm$, $y_\pm$, and $z_\pm$ are cartesian coordinates defining two $\R^3$ spaces with opposite orientations: right handed for $r^\prime>0$ and left-handed for $r^\prime<0$ (Reall, 2008). In terms of these coordinates, $$ds^2_{KN}\vert_{BL}\vert_{M=0, \ Q^2=0}=ds^2_K\vert_{BL}\vert_{M=0}=dt^2-(dx_\pm^2+dy_\pm^2+dz_\pm^2). \eqno{(38)}$$ In particular, from (30), $$ds^2_K\vert_{BL}\vert_{M=0}=dt^2-({{\Sigma}\over{\Delta}}{dr^\prime}^2+\Sigma d\theta^2+\Delta sin^2\theta d\phi^2).\eqno{(39)}$$ In terms of $(x^\prime,y^\prime,z^\prime)$, $r^\prime$ is given by $$r^\prime=\pm{{1}\over{\sqrt{2}}}\sqrt{(x_\pm^2+y_\pm^2+z_\pm^2-a^2)+\sqrt{(x_\pm^2+y_\pm^2+z_\pm^2-a^2)+4a^2z_\pm^2}}.\eqno{(40)}$$

\

Both the $KS$ and the $EC$ coordinate systems admit, at each $t$, the same foliations of the two $\R^3$ spaces:

\

i) Confocal ellipsoids of revolution $r^\prime=const.$, foci at $x_+^2+y_+^2=a^2$, $z_+=0$: $${{x_\pm^2+y_\pm^2}\over{{r^\prime}^2 +a^2}}+{{z_\pm^2}\over{{r^\prime}^2}}=1,\eqno{(41)}$$ with larger semi-axis =$\sqrt{{r^\prime}^2+a^2}$ and smaller semi-axis=$\vert r^\prime\vert$. For $r^\prime>0$, the ellipsoids corresponding to $r^\prime_\pm$ are the horizons $H_\pm$, embedded in the $(x_+,y_+,z_+)$ space.

\

ii) Confocal 1-sheet hyperboloids of revolution $\theta=const.$, foci at $x_+^2+y_+^2=a^2$, $z_+=0$: $${{x_\pm^2+y_\pm^2}\over{a^2sin^2\theta}}-{{z_\pm^2}\over{a^2cos^2\theta}}=1.\eqno{(42)}$$ The surfaces $\phi=const.$ for the $KS$ coordinates are discussed e.g. in Krasi\'nski and Pleba\'nski (2006), but for the $EC$ coordinates they are simply given by: 

\

iii) Planes through the $z_\pm$ axis.

\

Since the $EC$ system can accommodate the horizons $H_\pm$ and the ergosurfaces $S_\pm$ (see below), we shall restrict the discussion to this coordinate system. 

\

The {\it curvature singularity} of (30) is given by the condition $$\Sigma=0 \eqno {(43a)}$$ which, by (17), implies $$r^\prime=0, \ \theta={{\pi}\over{2}}.\eqno{(43b)}$$ By (36), $r^\prime=0$ defines the disks $D^2$ $$0\leq x_\pm^2+y_\pm^2=a^2sin^2\theta\leq a\eqno{(44)}$$ in the equatorial planes $z_\pm=0$. It is clear that its interiors ${\dot{D}}^2$ must be identified i.e. $(x_++\epsilon,y_++\epsilon)=(x_--\epsilon,y_--\epsilon)$ as $\epsilon\to\pm 0$ for those $x_\pm$, $y_\pm$ satisfying $x_\pm^2+y_\pm^2<a^2$, and that the boundary $$x_\pm^2+y_\pm^2=a^2 \eqno{(45)}$$ is the singularity. On one ``side" of ${\dot{D}}^2$ one has the $\R^2\times S^2$ region $r^\prime>0$ (with horizons and ergosurfaces), on the other ``side" one has another $\R^2\times S^2$ region which corresponds to $r^\prime<0$, but without horizons and ergosurfaces. (In the $\R^2$ factors, one $\R$ comes from $r^\prime>0$ and $r^\prime<0$, the other from the time coordinate.) It is easily seen that in the $r^\prime\to 0_\pm$ limit, the ellipsoids (41) degenerate into the disks (44), while the hyperboloids (42) degenerate into $(z_\pm=0$ planes $\setminus {\dot{D}}^2)$. 

\

The ergosurfaces $S_\pm$ (in the $r^\prime>0$ region) are determined by the zeros of $g_{tt}$. From (31), $$r^\prime_{S_{\pm}}(\theta)=M\pm\sqrt{M^2-(a^2cos^2\theta+Q^2)}. \eqno{(46)}$$ In particular $$r^\prime_{S_{\pm}}(0)=r^\prime_{S_{\pm}}(\pi)=r^\prime_\pm \eqno{(47)}$$ i.e. $S_\pm\equiv H_\pm$ at the ``north" and ``south" poles. Replacing (46) in (36) with $F(r^\prime)=0$, we obtain for both $S_+$ and $S_-$ the surfaces of revolution $${ {{x_+}_{S_\pm}(\theta)^2+{y_+}_{S_\pm}(\theta)^2}\over{(\sqrt{r^\prime_{S_\pm}(\theta)^2+a^2})^2} }+{ {{z_+}_{S_\pm}(\theta)^2}\over{r^\prime_{S_\pm}(\theta)^2} }=1\eqno{(48)}$$ which, together with horizons $H_\pm$, ellipsoids and hyperboloids corresponding to different sets of values of $M$, $a$, and $Q^2$ are plotted in the $y_+-z_+$ plane in Figures 2, 3 and 4. For $\phi={{\pi}\over{2}}$ ($y_+-z_+$ plane or $x_+=0$), $${y_+}_{S_\pm}({{\pi}\over{2}})=\sqrt{(r^\prime_{S_\pm}({{\pi}\over{2}}))^2+a^2}=\sqrt{(M\pm\sqrt{M^2-Q^2})^2+a^2}\eqno{(49)}$$ with $r^\prime_{S_+}({{\pi}\over{2}})=M+\sqrt{M^2-Q^2}>r^\prime_+>M$ and $r^\prime_{S_-}({{\pi}\over{2}})=M-\sqrt{M^2-Q^2}<r^\prime_-<M.$ So, for $Q^2=0$, $${y_+}_{S_+}({{\pi}\over{2}})=\sqrt{4M^2+a^2}, \ \ {y_+}_{S_-}({{\pi}\over{2}})=a,\eqno{(50)}$$ and, for $Q^2>0$, $${y_+}_{S_+}({{\pi}\over{2}})=\sqrt{(M+\sqrt{M^2-Q^2})^2+a^2}, \ \ a<{y_+}_{S_-}({{\pi}\over{2}})=$$
$$\sqrt{(r^\prime_{S_-}({{\pi}\over{2}}))^2+a^2}<\sqrt{{r^\prime_-}^2+a^2}<\sqrt{M^2+a^2}.\eqno{(51)}$$

\ 

The same structure of ellipsoids and hyperboloids are in the $y_--z_-$ plane; however, in this case, no of the ellipsoids corresponds to $H_+$ or $H_-$.

\begin{figure}[H] \centering
\includegraphics[width=.5\linewidth]{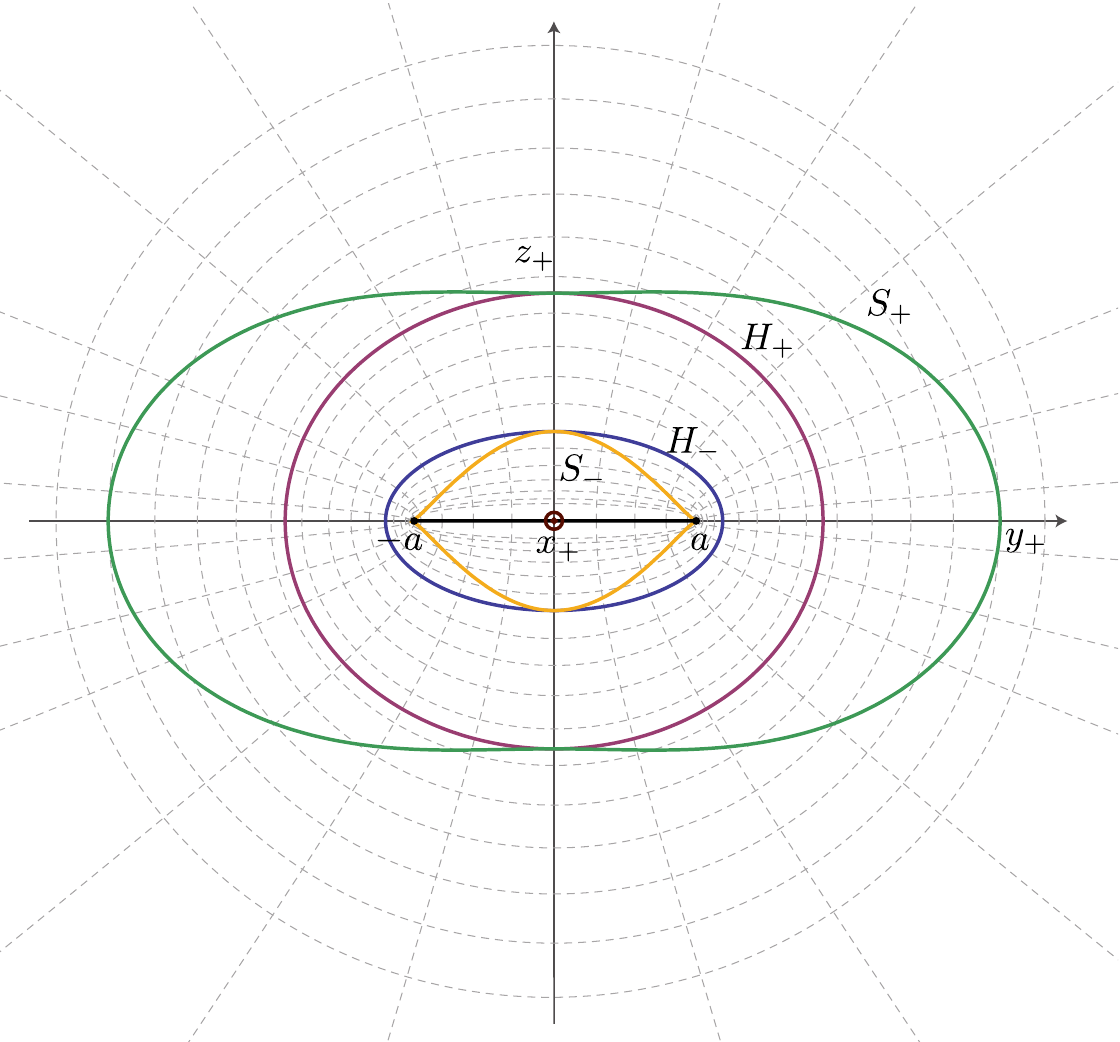}
\caption{Horizons, ergosurfaces, and ellipsoids and hyperboloids foliations in the $r^\prime>0$ region for the $K$ case. $H_\pm$ and $S_\pm$ for $a=.9$ and $M=1$.}
\end{figure}

\begin{figure}[H] \centering
\includegraphics[width=.5\linewidth]{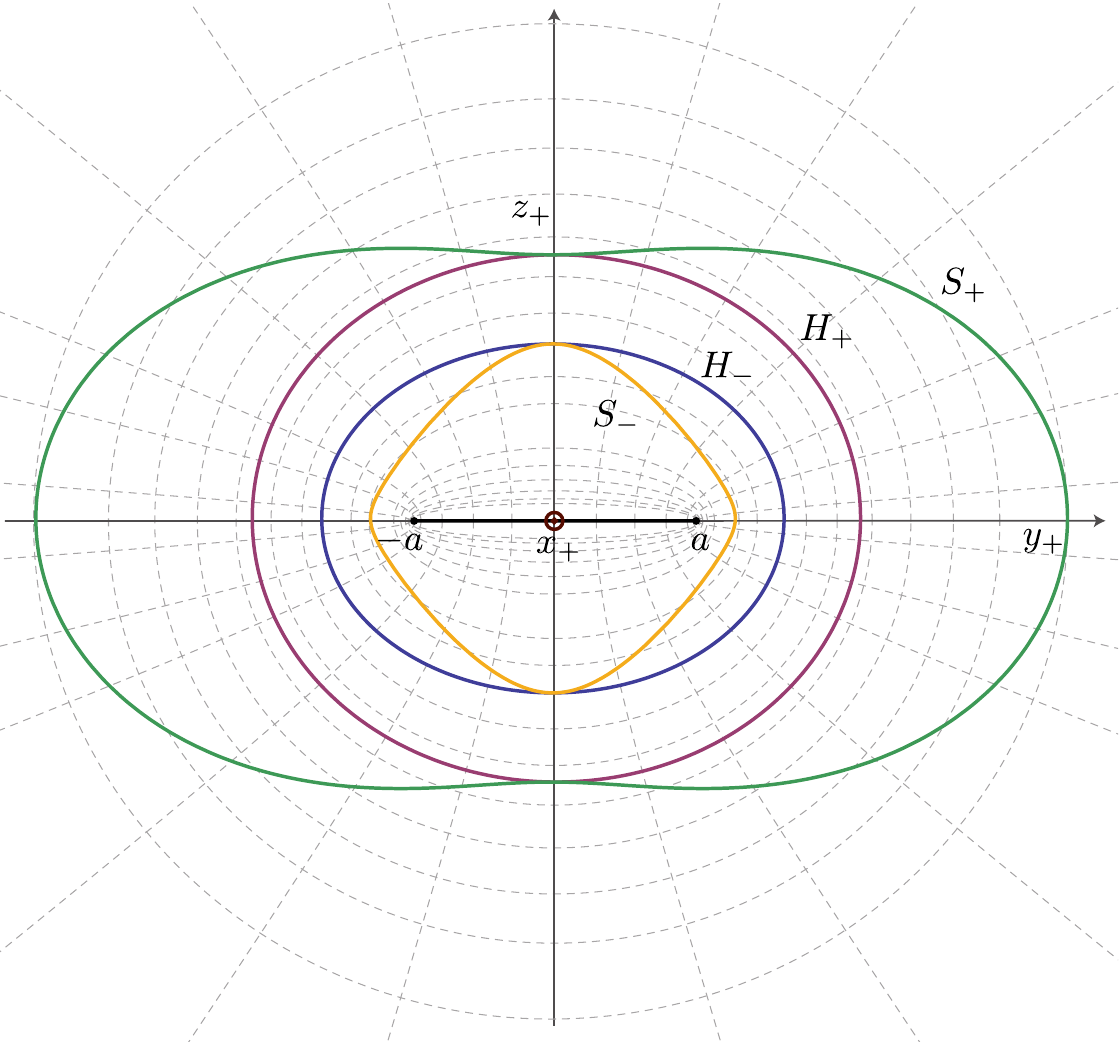}
\caption{Idem as Fig. 2 for the $KN$ case. $H_\pm$ and $S_\pm$ for $a=.9$, $M=1.3$ and $Q^2=.81$.}
\end{figure}

\begin{figure}[H] \centering
\includegraphics[width=.5\linewidth]{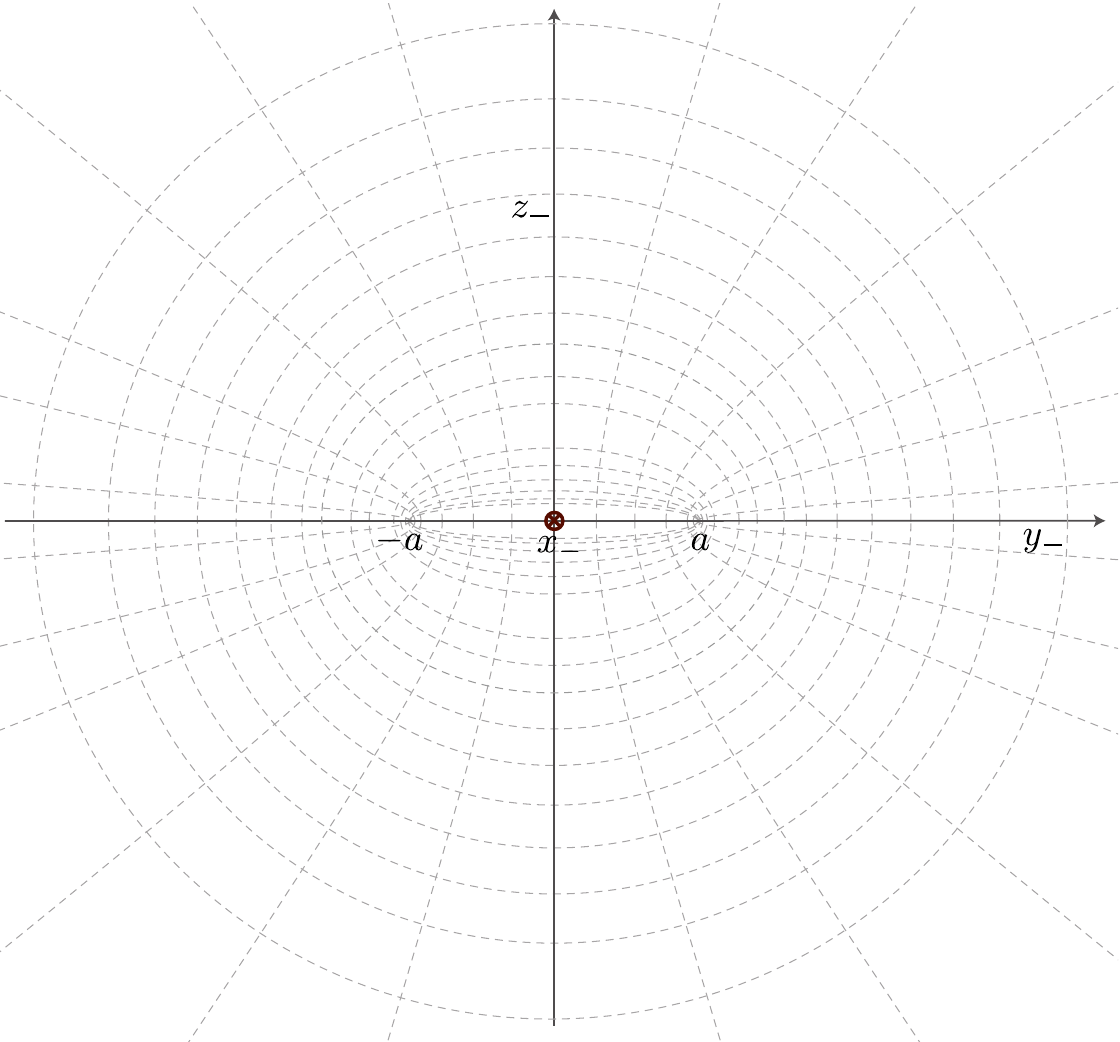}
\caption{Foliations with ellipsoids and hyperboloids of the region $r^\prime<0$ (plane $y_-/z_-$) for $a=.9$.}
\end{figure}
\

The spatial topology consists of two copies of $\R^3$ glued by an open disk of radius $a$ with its circular boundary being the singularity; each $\R^3$ is equivalent to $\{pt.\}\cup\R^+\times S^2\cong\{pt.\}\cup\R\times S^2$ where $pt.=(0,0,0)$ is the common origin. Only one of these $\R^3$'s contains the horizons and ergospheres. Including the time axis, one ends with two copies of $\{pt.\}\cup\R^2\times S^2$. The spatial topology can be formally ``viewed" by reducing the spatial dimension through the elimination of the $x_+$ and $x_-$axis: two $\R^2$'s joined at an open segment with the singularity being its end points (Figure 5). 
\begin{figure}[H] \centering
\includegraphics[width=.6\linewidth]{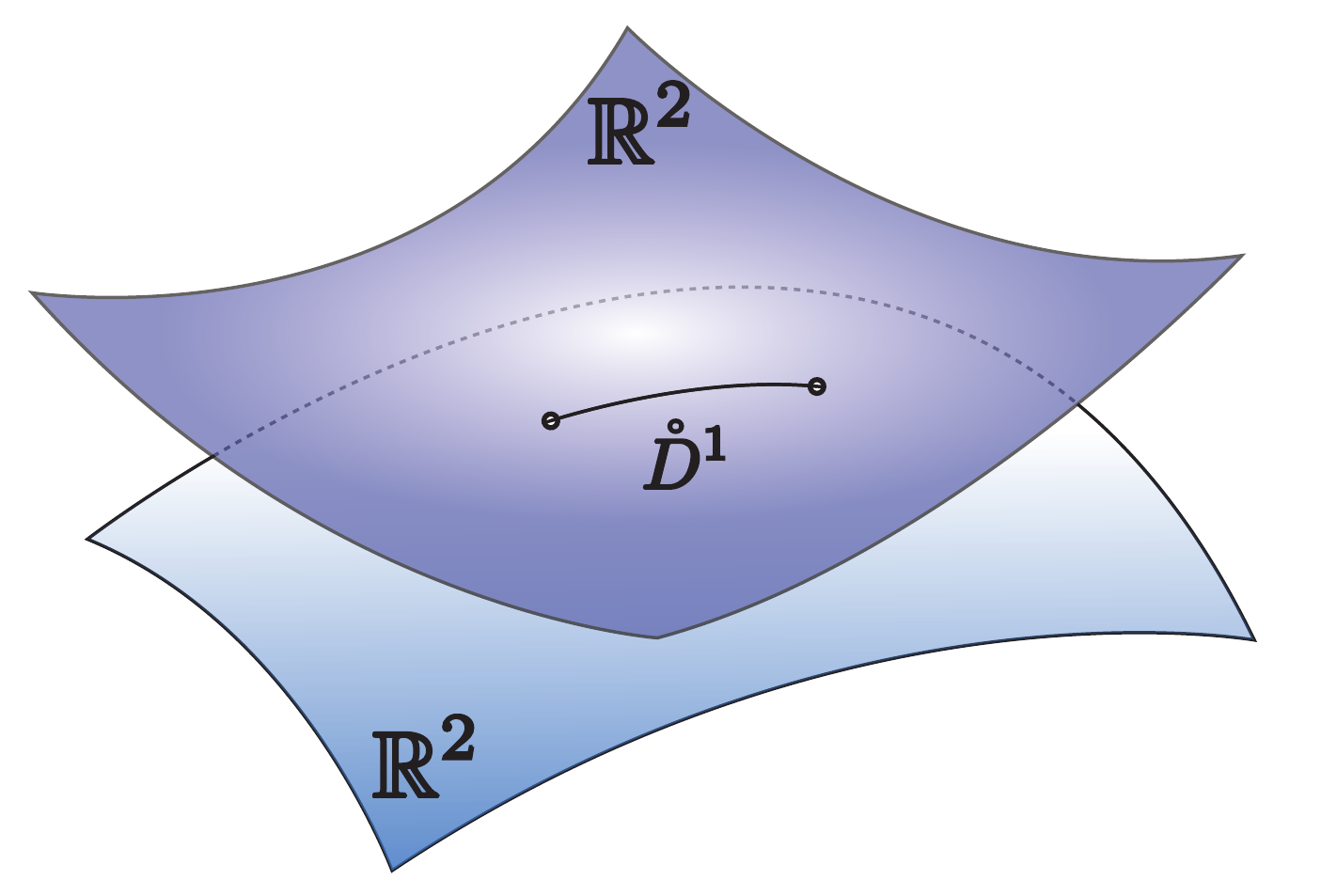}
\caption{Pictorial view of the spatial topology of the $K$ and/or $KN$ spacetimes with one spatial dimension ($x_\pm$-axis) eliminated.}
\end{figure}

{\bf 8. Penrose-Carter diagram}

\

For completeness, we present the Penrose-Carter diagram corresponding to the $K$ and $KN$ cases, consisting in the infinite ``vertical" repetition of the elementary cell illustrated in Figure 6. All elements: horizons and singularity are represented in the cell. The infinite tower is necessary to have a geodesically complete spacetime for geodesics that do not end at the singularity ring.

\begin{figure}[H] \centering
\includegraphics[width=.7\linewidth]{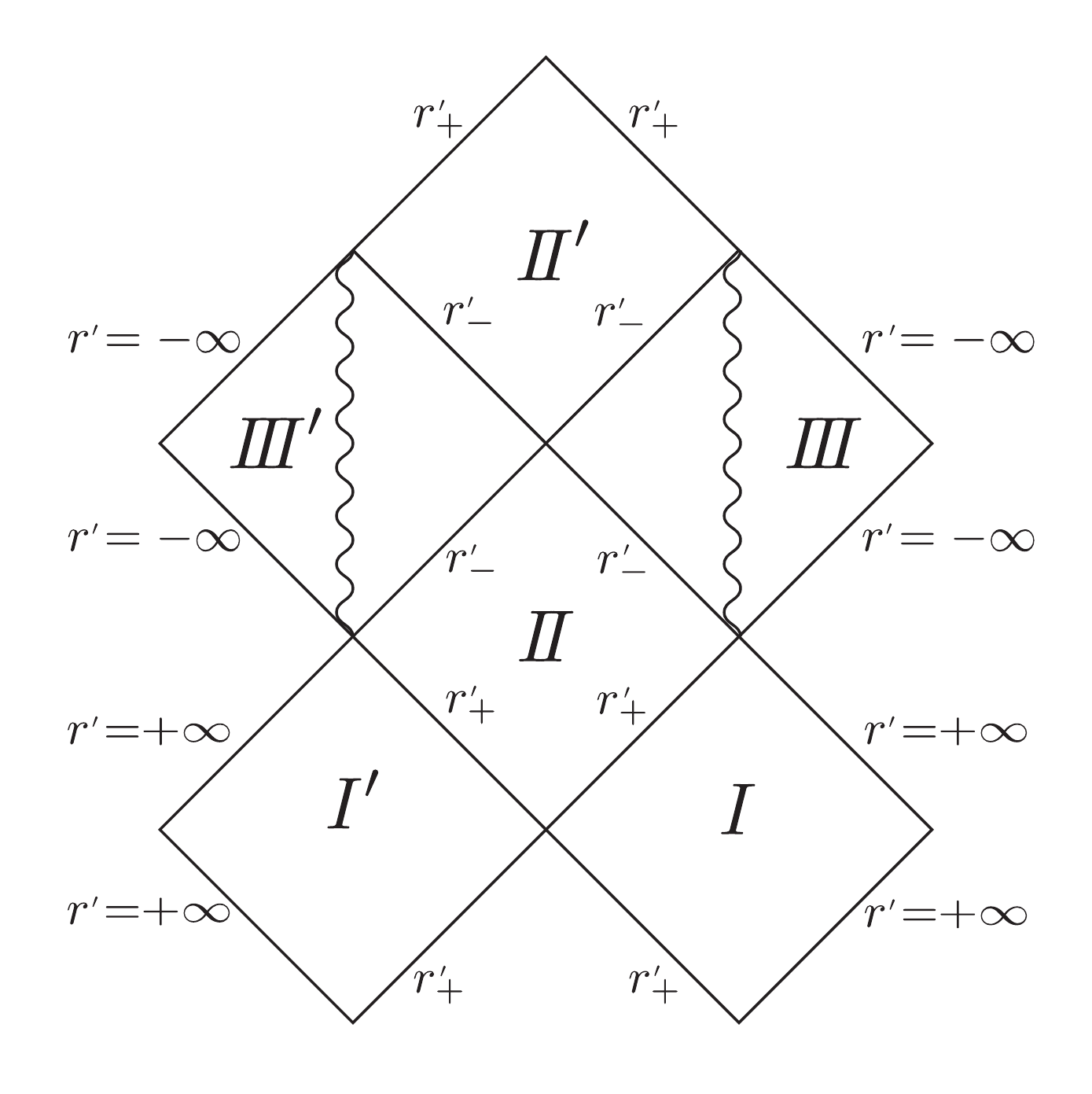}
\caption{ Elementary cell of the Penrose-Carter diagram of the $KN$  spacetime.}
\end{figure}

{\bf Acknowledgements}

\

This work was partially supported by the project PAPIIT IN105413, DGAPA, UNAM.

\

{\bf References}

\

1. Boyer, R. H. and Lindquist, R.W. {\it Maximal Analytic Extension of the Kerr Metric}. Journal of Mathematical Physics, {\bf 8} (1967), 265-281.

\

2. Carter, B. {\it Complete Analytic Extension of the Symmetry Axis of Kerr's Solution of Einstein's Equations}. Physical Review, {\bf 141} (1966), 1242-1247.

\

3. Carter, B. {\it Global Structure of the Kerr Family of Gravitational Fields}. Physical Review, {\bf 174} (1968), 1559-1571.

\

4. Drake, S. P. and Szekeres, P. {\it Uniqueness of the Newman-Janis Algorithm in Generating the Kerr-Newman Metric}. General Relativity and Gravitation, {\bf 32} (2000), 445-457.

\

5. Eddington, A. S. {\it A Comparison of Whitehead's and Einstein Formulas}. Nature, {\bf 113} (1924), 192. 

\

6. Finkelstein, D. {\it Past-Future Asymmetry of the Gravitational Field of a Point Particle}. Physical Review, {\bf 110} (1958), 965-967.

\

7. Kerr, R. P. {\it Gravitational Field of a Spinning Mass as an Example of Algebraically Special metrics}. Physical Review Letters, {\bf 11} (1963), 237-238.

\

8. Kerr, R. P. and Schild, A. {\it A New Class of Vacuum Solutions of the Einstein Field equations}. General Relativity and Gravitation, {\bf 41} (2009), 2485-2499. Original paper: R. P. Kerr and A. Schild, in: ``Atti del Convegno sulla Relativita Generale: Problemi dell' Energia e Onde Gravitazionali", G. Barb\`ere Editore, Firenze 1965, pp. 1-12.

\

9. Pleba\'nski, J. and Krasi\'nski. {\it An Introduction to General Relativity and Cosmology}, Cambridge University Press, Cambridge, (2006), p. 450. 

\

10. Newman, E. T. and Janis, A. I. {\it Note on the Kerr Spinning-Particle Metric}. Journal of Mathematical Physics, {\bf 6} (1965), 915-917.

\

11. Newman, E. T., Couch, E., Chinnapared, K., Exton, A., Prakash, A., and Torrence, R. {\it Metric of a Rotating, Charged Mass}. Journal of Mathematical Physics, {\bf 6} (1965), 918-919.

\

12. N\"{o}rdstrom, G. {\it On the Energy of the Gravitational Field in Einstein's Theory}. Proc. Kon. Ned. Akad. Wet., {\bf 20} (1918), 1238-1245.

\

13. Penrose, R. {\it Asymptotic Properties of Fields and Space-times}. Physical Review Letters, {\bf 10} (1963), 66-68.

\

14. Reall, H. {\it Black Holes}. Lecture Notes, University of Cambridge, (2008).

\

15. Reissner, H. {\it \"{U}ber die Eigengravitation des Elektrischen Feldes nach der Einsteinschen Theorie}. Annalen der Physik, {\bf 50} (1916), 106-120. 

\

16. Schwarzschild, K. \textit{\"{U}ber das Gravitationsfeld eines Massenpunktes nach der Einteinschen Theorie.Sitzungsberichte der Könichlich Akademie der Wissenschaften, \textbf{7}} (1916), 189-196.

\vspace*{\fill}
emails: brauer@ciencias.unam.mx, hugocm$_{-}$89@hotmail.com, socolovs@nucleares.unam.mx

\end{document}